# Pathological Analysis of Stress Urinary Incontinence in Females using Artificial Neural Networks

Mojtaba Barzegari[1], Bahman Vahidi[1*], Mohammad Reza Safarinejad[2], Marzieh Hashemipour[3]

1- Division of Biomedical Engineering, Department of Life Science Engineering, Faculty of New Sciences and Technologies, University of Tehran, Tehran, Iran.

2- Clinical Center for Urological Disease Diagnosis and Private Clinic Specialized in Urological and Andrological Genetics, Tehran, Iran.

3- Faculty of Computer Engineering, Amirkabir University of Technology, Tehran, Iran.

* P.O.B. 1439957131, Tehran, Iran, bahman.vahidi@ut.ac.ir

**Abstract**

**Objectives:** To mathematically investigate urethral pressure and influencing parameters of stress urinary incontinence (SUI) in women, with focus on the clinical aspects of the mathematical modeling.

**Method:** Several patients' data are extracted from UPP and urodynamic documents and their relation and affinities are modeled using an artificial neural network (ANN) model. The studied parameter is urethral pressure as a function of two variables: the age of the patient and the position in which the pressure was measured across the urethra (normalized length).

**Results:** The ANN-generated surface, showing the relation between the chosen parameters and the urethral pressure in the studied patients, is more efficient than the surface generated by conventional mathematical methods for clinical analysis, with multi-sample analysis being obtained. For example, in elderly people, there are many low-pressure zones throughout the urethra length, indicating that there is more incontinence in old age.

**Conclusion**: The predictions of urethral pressure made by the trained neural network model in relation to the studied effective parameters can be used to build a medical assistance system in order to help clinicians diagnose urinary incontinence problems more efficiently.

**Key words:** Urinary Tract, Urethral Pressure, Stress Urinary Incontinence, Artificial Neural Networks

**INTRODUCTION**

The diseases associated with the urinary and genital tract, generally termed as urology, are more prevalent in individuals older than 40 years in both, men and women, with the most critical subset of these problems being urinary incontinence.

Stress urinary incontinence (SUI) is very common among women and occurs as a result of a mechanical pressure like sneezing or jumping height. In this type of incontinence, any activity that increases the abdominal pressure (including laughing, coughing, sneezing, and straining) leads to urine leakage, which is due to urethral sphincter weakness. According to published reports in 2001, expenditure on urine incontinence treatments exceeded USD 16.3 billion. The reports indicate that 75% of the total costs were spent on the diagnosis and treatment of women [1].

Being more prevalent in women, SUI is mainly characterized by an increase in abdominal pressure in the absence of bladder contraction, which raises the vesical pressure to a level that exceeds urethral pressure, thus leading to an involuntary loss of urine. The abdominal pressure increases due to a mechanical incident like laughing, sneezing, jumping height, or any other tension in the body; this explains why SUI is considered a mechanical force.

Although the main cause of SUI remains unknown, a large number of physicians believe that SUI is caused by injuries to the pelvic floor neuro-musculature, which mainly occurs in women who have given birth vaginally. While not a life-threatening condition, SUI can have a detrimental impact on their quality of life.

Urodynamic testing is a group of detective methods used to evaluate urinary disorders, so as to protect the bladder and prevent urinary incontinence, and is mainly used to study the lower urinary tract. An important part of urodynamic testing documentation is the urethral pressure profile (UPP) diagrams, in which the pressure along the urethra is measured and plotted against the measurement location as the X axis (Figure 1). The UPP diagram has a high order of importance in clinical and pathological studies [2].

An artificial neural network (ANN) is a model used to process data, inspired by the neurological system of biological entities that processes information in a similar manner to the human brain. This system consists of a large number of connected computing elements, called neurons, which collaborate to solve a particular problem. An ANN learns new things by example, just like a human. An ANN is configured and trained to perform a specific task such as pattern recognition or information clustering. In biological systems, learning is accompanied by modifications in the synaptic connection, which is a similar function displayed by ANNs [3].

ANNs are the simplified models of neural structures as existing in the human brain. The ANNs approach computational tasks differently in comparison with the conventional computational methods and have the ability to study the governing rules by learning the input examples.

ANNs have different topologies and structures. In the manner of structure, there are two categories of ANN, namely feedforward and recurrent, and each category consists of many different topologies of ANNs. The known ANNs are multi-layer perceptron (MLP), radial basis function (RBF), Hopfield networks, and the Kohonen self-organizing map (KSOM) [3].

Each neural network has several computational neurons in its structure. These neurons have an activation function that computes the output of the neuron in regard to the inputs. The connections between the neurons have their own weights which are applied to the transmitted data. The learning stage of an ANN corrects these weights, using the input training set. This process is unique for each type of ANN. A typical MLP model is presented in Figure 2 [3].

As shown in Figure 2, the MLP model consists of several layers with different roles. Each layer includes several computational neurons. Each neuron has an activation function that compares the aggregates of input values with the bias value of the neuron and calculates the neuron output [3,4]. This operation is shown in Figure 3 schematically.

Over the past two decades, and especially in recent years, many advances have been made in informatics and medical engineering. Moreover, with the advent of modern computer science, new concepts such as artificial intelligence have entered man's life with critical effects on all aspects of his life. But, owing to the importance of the health field, the use of artificial intelligence and computational techniques in medicine is more significant, leading to numerous advances in terms of disease prevention and diagnosis by the use of computer concepts [5,6]. One of the impacts of these medical advancements is the introduction of decision support systems (DSSs) in medicine. Intelligent medical DSS is a commonly used term referring to intelligent systems designed to provide physicians and other health professionals with clinical decision support [7]. By using the decision-making systems, it is possible to make logical and knowledge-based decisions without being affected by environmental conditions or human errors which may occur, for instance, due to fatigue. Thus, DSSs lead to an increased accuracy and improvement of diagnostic quality in healthcare [7].

In modern medicine, artificial intelligence contributes to the creation and application of medical knowledge while being widely used to offer warnings, reminders, and confirmations of diagnostic decisions. The common advantages of artificial intelligence include the emergence of DSSs that have numerous applications in medical sciences [8]. There is a strong relationship and alignment between knowledge and data, indicating that the discovery, diagnosis, interpretation, and treatment of diseases are affected by the related clinical data. This degree of communication and alignment differs depending on the problem and the issues involved. The intelligent computing models make decisions based on data rather than knowledge, and these have become more successful today [9]. In recent years, ANNs, a branch of artificial intelligence, have been suggested as an alternative tool for the classification of diseases.

There are different diagnostic methods for medical diagnosis that can be used as DSSs in medicine. Common diagnostic approaches require the formulation of rules and instructions that are able to systematically regulate the relationships between data while they analyze the input data, based on which accurate diagnoses are obtained. However, in most cases, it is very difficult to formulate these rules and instructions, particularly when it comes to large data sets. Special techniques, such as pattern classification which is among the main mechanisms of ANNs, are used as alternatives to the rule-based systems [10,11]. The ANNs, considered as one of these important methods in analyzing data, are much more effective than other methods in medical diagnosis. Medical diagnostics enjoy the benefits of neural networks that are considered a method of categorizing diseases. The neural networks have the ability to learn without requiring the knowledge of underlying rules, allowing them to process new data successfully. Therefore, the use of neural networks is more useful and effective than traditional diagnostic methods. Furthermore, the ANNs act as a powerful tool to help clinicians in analyzing, modeling, and finding relationships between medical data for which the normal methods, such as the rule-based and statistical systems, are not appropriate and in many cases respond incorrectly. For example, patients may not have the same early symptoms while having similar reactions. In these cases, traditional systems such as axial systems act incorrectly [7,8,11-13].

The medical DSSs prevent potential mistakes that could be caused by fatigue or inexperienced clinical professionals. These systems are also able to analyze medical data in a shorter time and greater detail, while delivering the extracted knowledge to the clinical specialist in a simpler and faster manner. Computer-based DSSs can also be set up on the World Wide Web for online access, thus allowing them to work without interruption [11,13-15].

Today, early diagnosis of many diseases is crucial for their treatment. The processes of disease diagnosis and categorization could benefit from modern computer technologies, such as DSSs. The superiority of neural network techniques over regular programming is based on their problem-solving ability for problems which either have current solutions that are very complex or do not have algorithmic solutions [16]. These networks are modern, desirable tools that can help clinicians diagnose and decide on treatment effectively.

Regarding urinary tract and genital diseases, it is possible to analyze clinical data using neural networks. For example, prostate cancer as associated with the genital tract has great importance because of its high prevalence, and an artificial intelligence to examine this topic has been studied in several investigations [17].

Detecting prostate cancer by conventional methods involves utilization of simple or single-marker analyses, referring to a limited number of clinical markers, which could lead to incorrect diagnosis and diagnostic errors. In terms of etiology, accurate analysis of prostate cancer often requires a number of variables such as age, race, biochemical and clinical markers, volumetric prostate analysis, and morphological changes. Human beings are not capable of analyzing the complex and nonlinear relationships between these variables without making errors [18]. As a result, it is necessary to have a diagnostic tool with the capability of analyzing several parameters while interpreting their interactions, thus achieving coherent and accurate results through a multivariate analysis.

Apart from limitations of the common methods for diagnosis of prostate cancer, the advancement of modern computer-aided methods has encouraged the need for decision-making tools that can use specific disease parameters to help patients and clinicians make better-informed and more accurate decisions [15].

The use of clinical DSSs as high-precision tools can help diagnose cancers and reduce their diagnostic errors. In other words, developing these algorithms will help clinicians diagnose diseases, particularly cancer, while improving diagnostic outcomes. One such method is a set of advanced computer techniques that increase the accuracy of prostate cancer diagnosis, whereas the use of other non-invasive or even invasive methods for detecting this type of cancer can lead to errors [19-21].

In this research, the urethral pressure is investigated by using the UPP. This investigation has been performed by extracting several patients' data from UPP and modeling their relations and affinities. The urodynamic data of several patients was digitized and then analyzed by an ANN model in order to study the effective parameters on urethral pressure and compare the results between patients. The main contribution of this study is to focus on the clinical aspects of ANN results instead of having a pure analytical review, which is commonly found in similar studies. Another unique aspect of this study is that the urethral pressure with regard to the chosen effective parameters is studied along the entire length of the urethra, instead of being evaluated only at certain points.

**MATERIALS AND METHODS**

The effective factors behind a biological phenomenon are not limited to the known parameters. A vast plethora of reasons are involved in the creation of one sample biological phenomenon. In most cases and due to its complexity, the researchers cannot identify all the important factors, even in this modern technological era. To prove the last statement, we may mention the fact that if humankind had complete knowledge about biology, the full control over biological disorders like HIV, cancer, etc. would have been accomplished. However, the lack of complete information in this field is apparent through the many challenges encountered in the study of the human body, which is a complex system. The urinary tract is one such human organ whose biological phenomena and disorders are not completely understood by the researchers studying them.

One of the studied issues of the urinary tract is urethral pressure. There are several parameters which are responsible for urethral pressure. The most significant parameter is the distance from the bladder and support muscle—in other words, the pressure gradient through the urethra. From an engineering perspective, the urethral pressure depends on the spot in which the pressure is measured. This fact may be noticed in the UPP plots (horizontal axis indicates the location of the probe that measures the pressure). Thus, the first important parameter in urethral pressure is the distance from the bladder.

If the position was the only parameter that affected urethral pressure, the UPP plots would be identical for all patients, but UPP plots are unique for each patient. There are several other parameters which should be analyzed during the study of urethral pressure. For instance, bladder volume, urinary tract geometry, and even the drugs used by patients seem to be other effective parameters for UPP plot variations. Hence, the second parameter considered in this research is the patient's age. It should be noted that other parameters are neglected in order to simplify the ANN model.

The data obtained from the Hashemi Nejad Hospital include complete urodynamic testing for several patients. The most important part of this documentation relates to UPP, which should be converted to computer data that can be analyzed. Urethral pressure as a function of position in the urethra is obtained from these UPP plots for each patient. As mentioned earlier, the next important parameter required for analysis is the patient's age which is calculated based on the test date and patient's birthday. In Figure 4, the overall procedure has been illustrated. The last step in this diagram is the analysis of this digitized data, in order to study the effect of the selected parameters (distance and age) on urethral pressure. To accomplish this task, two approaches may be followed: 1. conventional interpolation and 2. neural networks.

In the first approach, conventional interpolation, a surface is interpolated based on the distance and age data of all studied patients. The urethral pressure could be evaluated at each point of this surface, depending on the relevant age and distance. Despite being well suited for several linear and simple nonlinear problems, this approach is not applicable for data with a high order of nonlinearity or noisy data. In addition, it may be necessary to extrapolate the data.

Most of the time, a high order of nonlinearity and noisy behavior may exist in the curves obtained from the clinical data. Since the conventional interpolation methods cannot cope with this kind of data, another method should be utilized. As mentioned in Karray's book [3], practical curve fitting may be obtained based on the neural network techniques, in the presence of noisy and nonlinear data. In addition, in the neural network methods, interpolation and extrapolation phases are handled simultaneously. Trained ANNs are able to model the mutual effects between parameters according to the training data. In addition, the noisy data is implicitly filtered during the training phase. The accuracy of predictions made by neural networks depends on the input data; hence, it is important to provide a rich set of data (more variant data) during training.

In this study, an MLP neural network model has been utilized to approximate the relation between the position in the urethra, the age of the patient, and urethral pressure. This type of ANN, along with RBF networks, is widely used for function approximation purposes.

The MLP networks typically consist of three or more layers (Figure 2). The first layer, called the input layer, contains the same number of neurons as the input parameters—for example, two neurons for age and distance, respectively. The input layer neurons have a linear activation function, which yields that inputs are transferred to outputs without any changes. The second layer is a hidden layer which is connected to the outputs of the input layer. The computational power of an MLP ANN depends on the neurons available in the hidden layer. Normally, only one hidden layer is utilized in the MLP ANNs; however, this number may be increased to two or more layers. The increase in the neuron number in the hidden layer leads to the increase in computational intelligence and network learning power. The activation function of the hidden-layer neuron plays a significant role in the neural network performance and accuracy. Sigmoid, linear, hyperbolic tangent, and binary signum are typical activation functions used in the hidden layer. In this research, the hyperbolic tangent is utilized for the hidden-layer neurons. The last layer in the MLP network is the output layer which is connected to the outputs of the hidden layer. This layer generates the results of the MLP network. The number of neurons in this layer is equal to the number of target values—for example, one for the urethral pressures in this research. Similar to the input layer, the activation function of neurons in this layer is also linear. Figure 5 represents the ANN model of the current research.

In most cases, the input data are normalized prior to the training phase, which leads to a better performance of the neural network model. Thus, the distance parameter (position in the urethra) is normalized to a value between zero and one. The age parameter varies from 20 to 76, which yields a small variation. So, normalization is not required for this parameter.

The training set contains 6453 data points, which include the normalized position in the urethra, the age of the patient, and the urethral pressure expressed in cmH2O. This input set is analyzed using both conventional interpolation and neural network models, in order to compare these approaches. The models are implemented in the MATLAB software package.

**RESULTS AND DISCUSSION**

The urodynamic data of UPP diagrams in patients with SUI, as obtained from the Hashemi Nejad Hospital in Tehran, is plotted in three dimensions. One axis to display the position in the urethra (distance from the bladder), another axis to display the age of patients, and a third to display pressure in cmH2O are considered. In fact, in the direction of the age axis, there are 22 unique UPP diagrams which are prepared from patients' information of all ages. Figure 6 displays the resulting three-dimensional plot.

In the next step, a surface is plotted using interpolation in order to demonstrate the inefficiency of conventional techniques in comparison with the methods used in this study. The plotted surface is shown in Figure 7.

The surface is used to plot the 2D contour showing the results of the conventional method. If we look from the top of the surface and display the higher level points (points with higher pressure) by red and the lower level points by blue and the other levels by a relevant color between these two colors, a contour would be generated. This contour illustrates the urethral pressure as a function of distance from bladder and age, which has been shown in Figure 8. As can be seen, this contour is very erratic and cannot be analyzed.

After training the neural network and modeling the clinical data, the network was prompted to generate a surface similar to the interpolated one. This new surface is plotted in green alongside the interpolated plot (red) in Figure 9. Now, if we turn the surface (the one that is generated by neural network model) to a contour, an arranged and analyzable contour would be obtained as given in Figure 10.

Figure 6 shows the points which have been used as inputs for the neural network training. These points are extracted from UPP diagrams and belong to 22 patients diagnosed with stress urinary incontinence. As is distinctive in the figure, distinguished curves are related to individual patients with different ages and are completely detectable across the age axis.

The first attribute in Figure 6 is a low amount of data presented in the age parameter. As indicated by the points located in the three-dimensional space, the distance from bladder parameter (position in the urethra), which is displayed by the characteristic length after normalization, has enough data for every patient at a specific age (as noted above, these data are extracted from the UPP diagrams of the patients). But for the age parameter, there is only distinct data for 22 cases, which is the number of studied patients. So, insufficient data examined age parameter is one of the shortcomings of the presented model, caused by a relatively low number of patients being studied. The total number of training data—that is, the number of points which are available in this figure is 6453 points.

Figures 7 and 8 represent the plotted surface by using the interpolation method and the generated contour using the surface, respectively. This surface is very irregular and therefore the resulting contour is chaotic too. The obtained contour is not clearly analyzable for discussion. This chaotic result is because of the conventional interpolation method used to plot the surface, which principally has a low performance in dealing with data possessing a high level of nonlinearity or noise. In fact, the plotted surface passes fully through all the data points as applied by the interpolation procedure. As these points are related to different patients at different ages, an explicit relation between the extracted data from each patient is not expected to be observed, and the plotted procedure and its related contour also reflect this disaffiliation. In fact, it is difficult to conceive a relationship between the pressure in the urethra with both the extracted parameters (position in the urethra and age), and the interpolated contour and surface confirm this claim. However, a trend for the increase and decrease of urethral pressure as a function of the two studied parameters could be found, which was achieved by using ANNs.

In Figure 9, the surface obtained from the trained neural network is displayed. Although at first it seems that the surface does not have a significant correlation with the extracted clinical data, generating the derived contour does illustrate the achievement of the desired results (Figure 10). Comparing the contour with the clinical data (which are used as neural network training input) and the contour generated by the interpolation functions (Figure 8) shows the efficiency of the contour as plotted by the neural network in demonstrating the trend of data and the impact of the studied parameters of urethral pressure. As mentioned, this process is limited to a small sample of clinical data (only 22 patients). So in order to achieve a more accurate and superior model, the first step is to increase the number of patients, which will then provide an increase in the parameter of age-related data.

The contour in Figure 10 demonstrates that in the examined patients, the maximum pressure occurs in the middle of the urethra. This area is roughly located where the urethral sphincter is placed [22-24] and this matter represents the accuracy of predictions made by utilizing the trained neural network (it could not be analyzed from the generated contour by interpolation). By increasing the patients' age up to about 58 years, the peak urethral pressure increases; however, the period beyond 58 years is associated with a decrement. The high-pressure zone is wider at young ages (i.e. only one high-pressure point and the other regions have relatively low pressure). In older ages, lots of low-pressure zones are observed across the urethra length, which is a strong reason for more incontinence in these ages [22-24].

Despite the small number of studied patients, the results of the trained neural network indicate a good relationship between the network output and urological analysis. By reinforcing the patients' database and making it available for the neural network model, we can achieve valuable results in diagnosing conditions that have not been examined in urodynamic tests.

**CONCLUSION**

In the current study, by using information and diagrams pertaining to a urethral pressure profile and extracting them from the urodynamic tests of several patients, an ANN was trained that could predict the trend of urethral pressure as a function of two select influencing parameters (distance from bladder and age). The following conclusions are obtained from the results:

- From the predictions of the trained neural network in relation to the effect of age and the position in which the pressure was measured across the urethra (normalized length) on urethral pressure, it can be concluded that the ANNs have great potential to anticipate the governing relations in biological phenomena.

- The prediction made by the neural network indicates that among the examined patients (22 cases), the highest pressure occurs in the middle of the urethra length, which is mostly located in the area where the urethral sphincter is placed; this finding acts as a confirmation of the trained neural network's ability to predict accurately.

- In the elderly, there is an increase in the number of low-pressure zones throughout the urethra length, indicating that there is more incontinence in older ages.

In future studies, it is suggested to increase the sample size of patients in order to achieve better accuracy in quantitative predictions made by the ANN model. It is also worth mentioning that extracting additional parameters that can affect urethral pressure and updating the neural network model to include them will result in an expert system that is better equipped to make precise predictions.

**Figure legends**

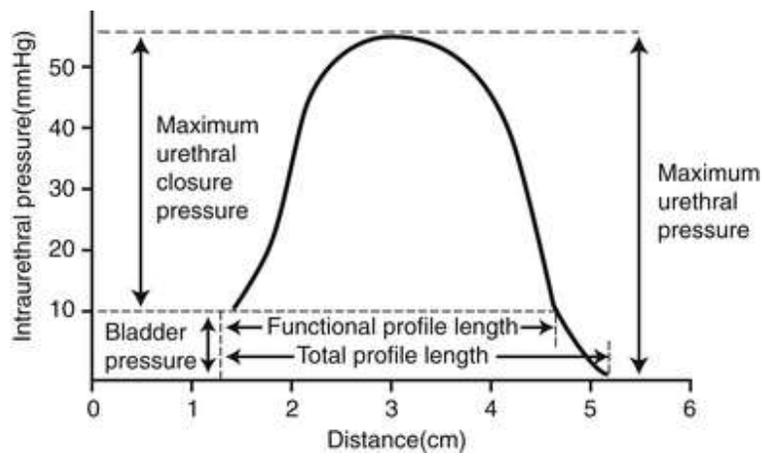

**Figure 1**. Urethral Pressure Profile UPP

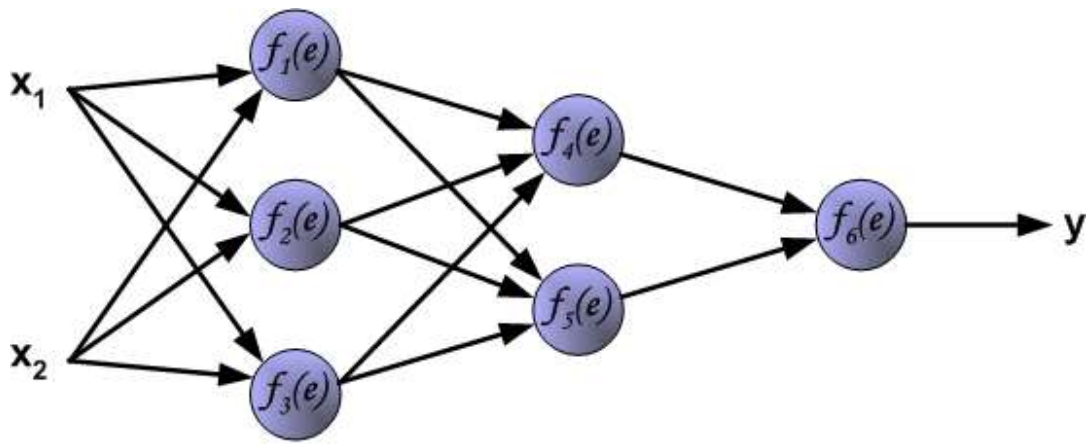

**Figure 2**. Simple Multilayer Perceptron Network (MLP)

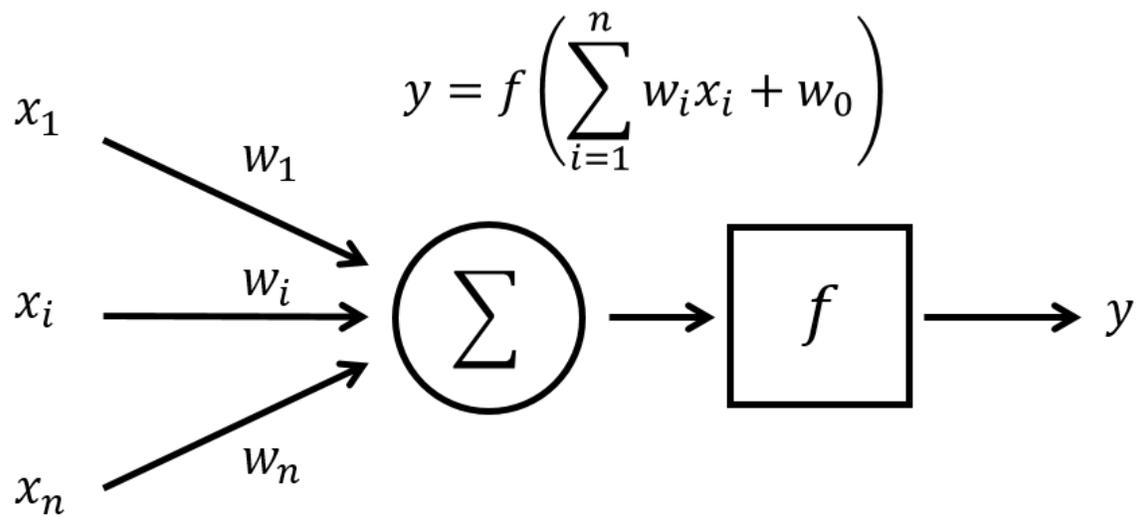

**Figure 3**. Mathematical model of a computational neuron

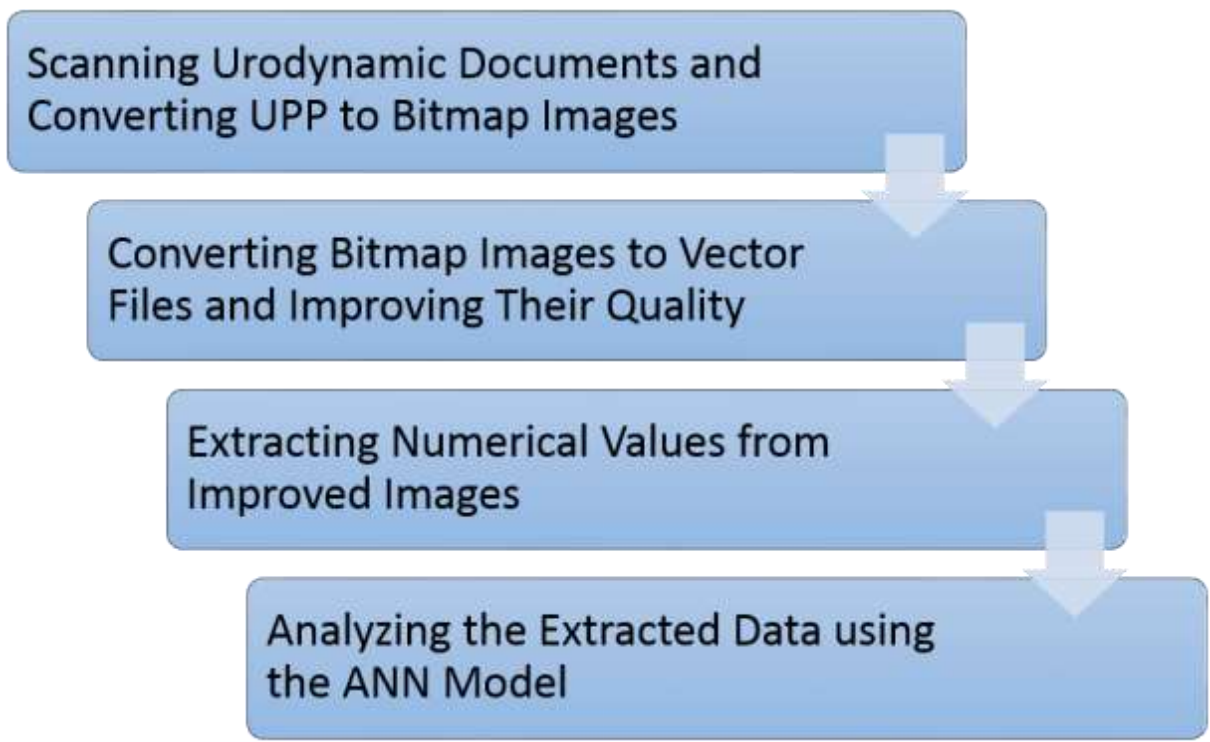

**Figure 4**. Overall procedure of present study

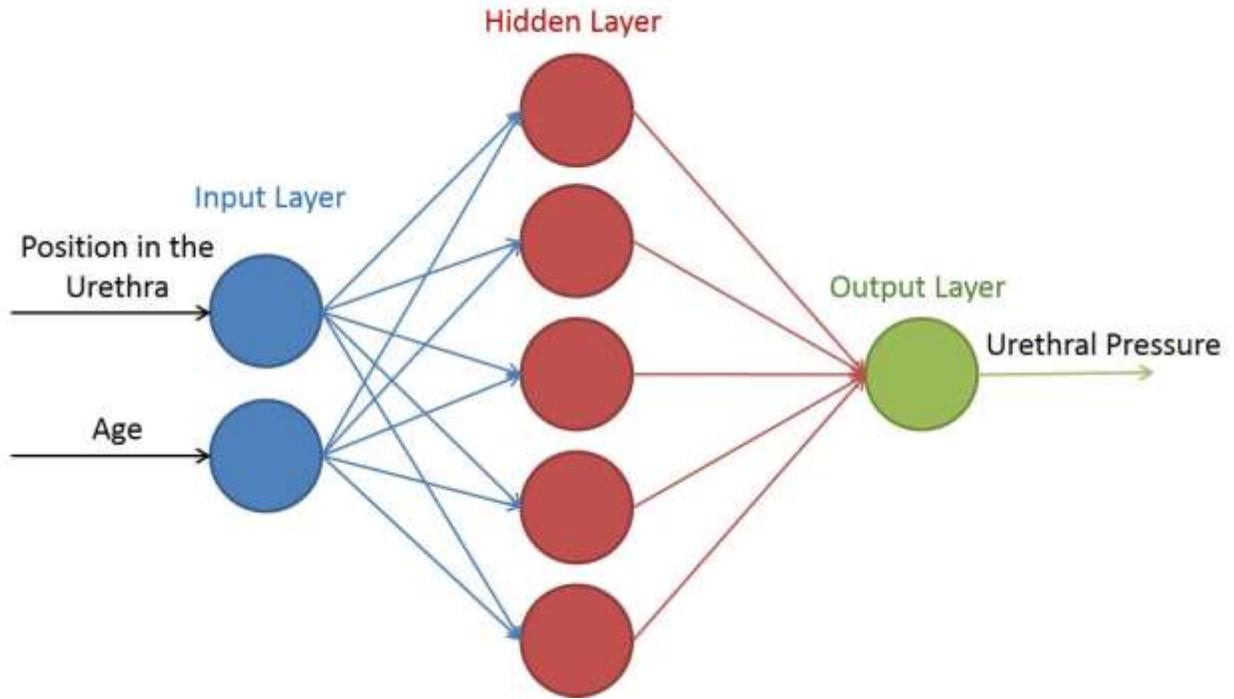

**Figure 5**. Multilayer Perceptron (MLP) network model developed in present study

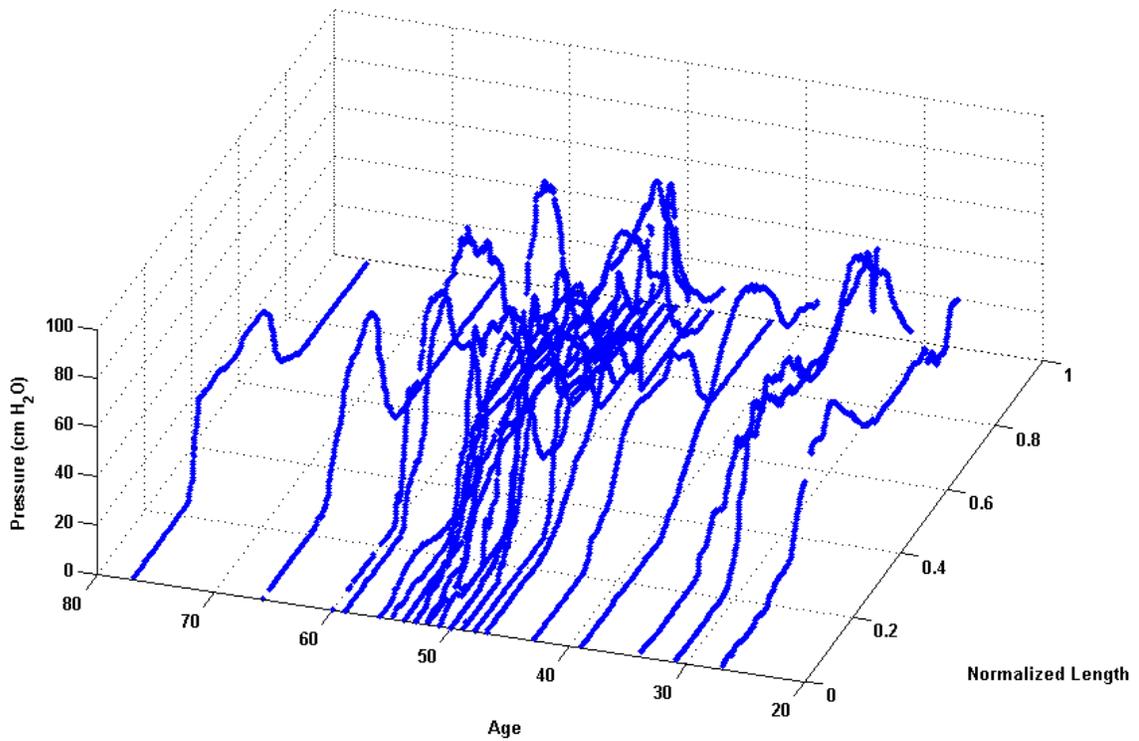

**Figure 6**. Data collected from patients plotted in 3D axis

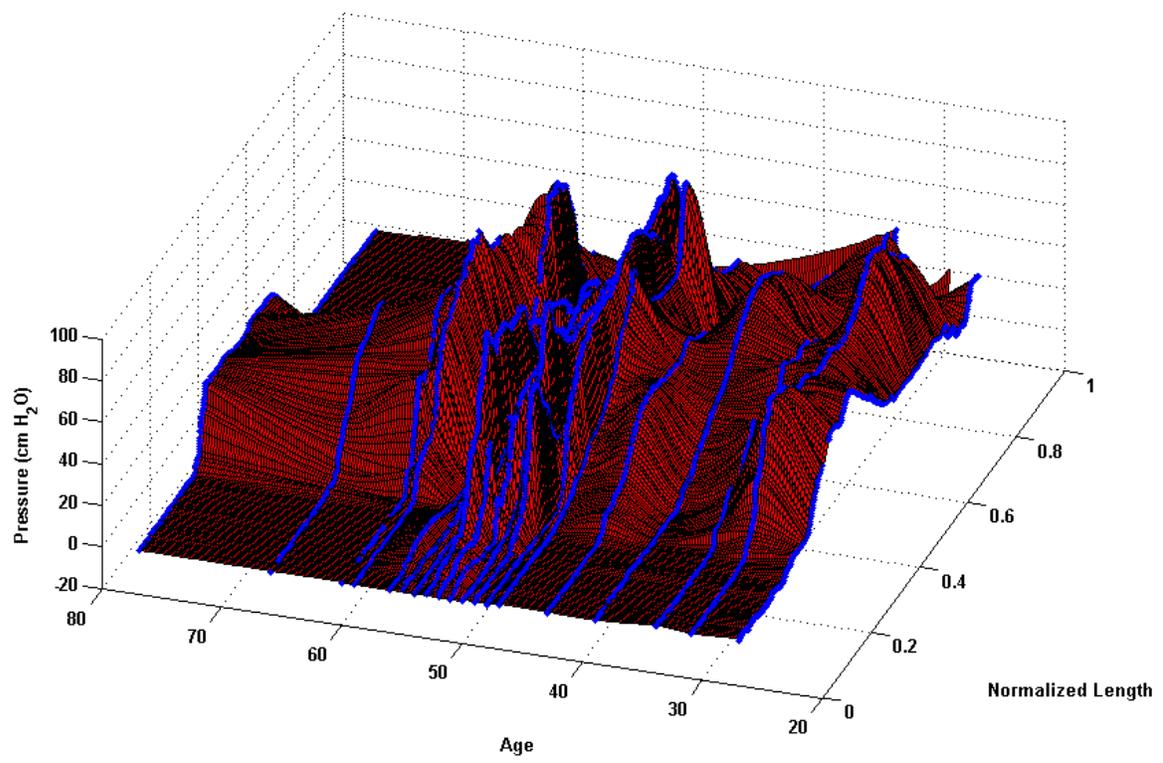

**Figure 7**. Generated surface by using interpolation methods

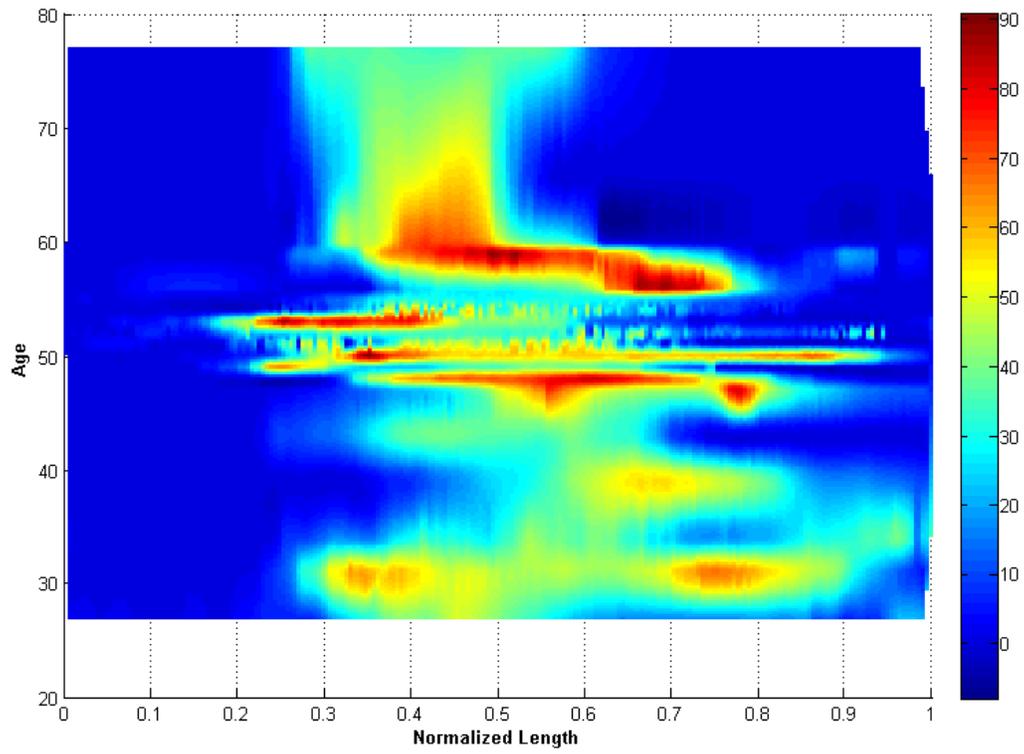

**Figure 8**. Contour of pressure variations as function of position in the urethra and age by using interpolation methods

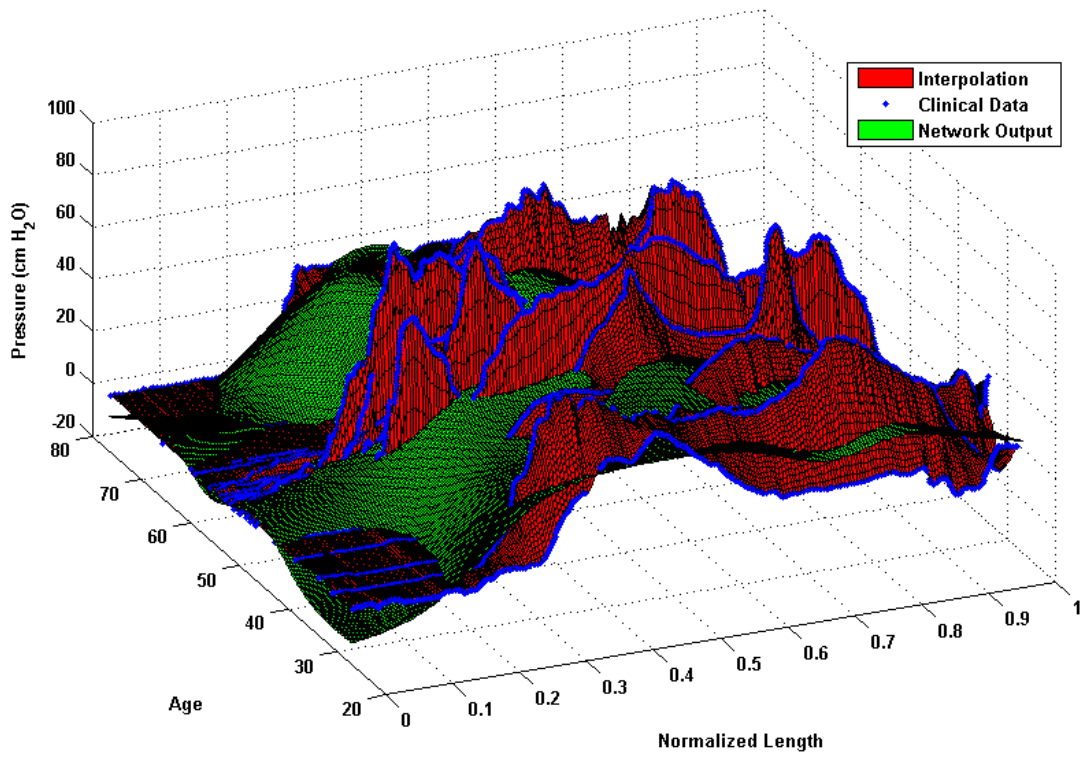

**Figure 9**. Comparison of generated surface using interpolation method (red) and ANN model (green)

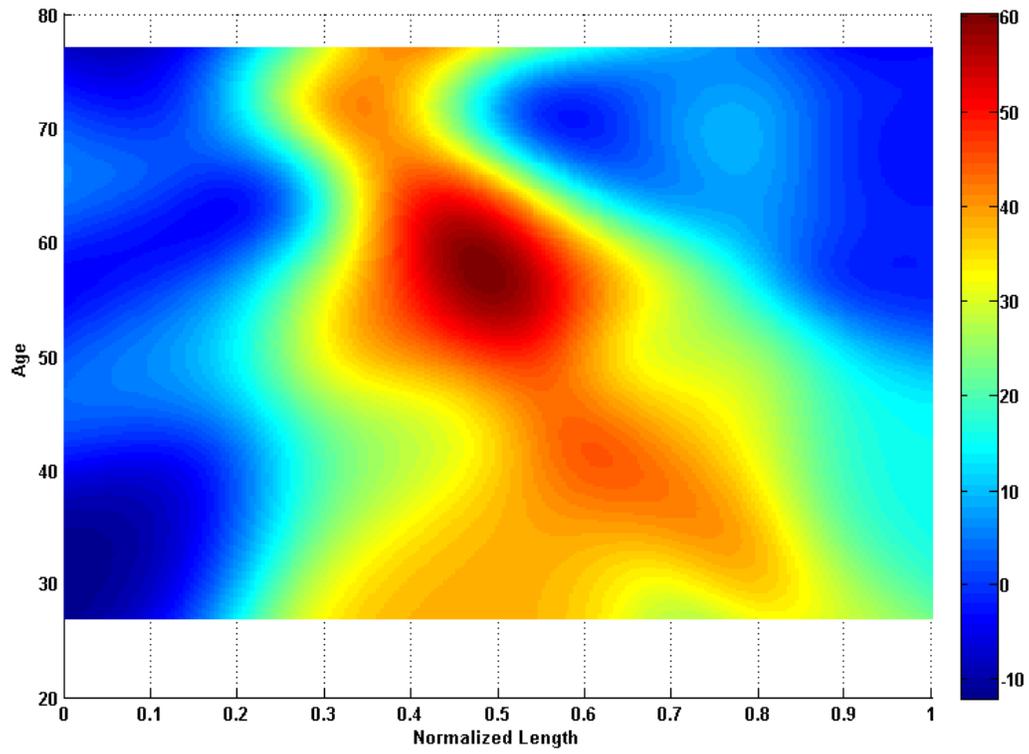

**Figure 10**. Contour of pressure variations as function of position in the urethra and age by using the ANN model developed in present study